\documentclass[twocolumn,preprint,aps,prb,citeautoscript,draftams,
math,amssymb,showpacs,superscriptaddress,floatfix]{revtex4-1}
\usepackage{graphicx}
\usepackage{dcolumn}
\usepackage{bm}
\usepackage[T1]{fontenc}
\usepackage{lmodern}
\usepackage{natbib}
\usepackage{makecell}
\usepackage{booktabs}
\usepackage{multirow}

\usepackage[flushleft]{threeparttable}

\usepackage{amsmath,amssymb}
\usepackage{graphicx,textcomp}
\usepackage{subfigure}
\usepackage{color}
\usepackage[colorlinks=true,
            linkcolor=blue,
            urlcolor=blue,
            citecolor=blue]{hyperref}

\begin{document}

\title{The role of N defects in paramagnetic CrN at finite temperatures from first-principles}

\author{E. Mozafari}
\email[Electronic mail:] {elhmo@ifm.liu.se}
\affiliation{Department of Physics, Chemistry and Biology, Link\"{o}ping University, SE-58183 Link\"{o}ping, Sweden}

\author{B. Alling}
\affiliation{Department of Physics, Chemistry and Biology, Link\"{o}ping University, SE-58183 Link\"{o}ping, Sweden}

\author{P. Steneteg}
\affiliation{Department of Physics, Chemistry and Biology, Link\"{o}ping University, SE-58183 Link\"{o}ping, Sweden}

\author{Igor A. Abrikosov}
\affiliation{Department of Physics, Chemistry and Biology, Link\"{o}ping University, SE-58183 Link\"{o}ping, Sweden}
\affiliation{Materials Modeling and Development Laboratory, NUST "MISIS", 119049 Moscow, Russia}
\affiliation{LOCOMAS Laboratory, Tomsk State University, 634050 Tomsk, Russia}

\date{\today}

\begin{abstract}
Simulations of defects in paramagnetic materials at high temperature constitute a formidable challenge to solid state theory due to the interaction of magnetic disorder, vibrations, and structural relaxations. CrN is a material where these effects are particularly large due to a strong magneto-lattice coupling and a tendency for deviations from the nominal 1:1 stoichiometry. In this work we present a first-principles study of nitrogen vacancies and nitrogen interstitals in CrN at elevated temperature. We report on formation energetics, the geometry of interstital nitrogen dimers, and impact on the electronic structure caused by the defects.  We find a vacancy formation energy of 2.28 eV with a small effect of temperature, a formation energy for N interstitial in the form of a $\big<111\big>$ oriented split-bond of 3.77 eV with an increase to 3.97 at 1000 K. Vacancies are found to add three electrons while split bond interstitial adds one electron to the conduction band. The band gap of defect-free CrN is smeared out due to vibrations, though it is difficult to draw conclusion about the exact temperature at which the band gap closes from our calculations. However, it is clear that at 900 K there is a non-zero density of electronic states at the Fermi level. At 300 K our results indicate a boarder case were the band-gap is about to close.
\end{abstract}
\maketitle
\section{Introduction}
\label{1}
Transition metal nitrides (TMNs) belong to the hard refractory materials which typically crystallize in the rocksalt structure. 
They are widely used as hard protective coatings on cutting tools and for coatings in metal-forming or plastic-moulding applications\cite{Vetter1995,Persson2001}. Amongst these, chromium nitride has in addition to its practical applications been of interest due to its fascinating magnetic, optical and electronic properties. It is well known that CrN is paramagnetic with a $B1$ structure at room temperature while at N\'{e}el temperature ($T_{N}$), in the range between 270-286 $K$, undergoes a phase transition to antiferromagnetic (AFM) with orthorhombic structure\cite{Corliss1960,Rivadulla2009,Wang2012}. Theoretically the magnetic stress is thought to be the driving force for the lattice distorsions\cite{Filippetti1999,Filippetti2000}. Thus the magnetic transition at temperatures around 280 $K$ is associated with the structural transition\cite{Corliss1960}. On the other hand, no sign of magnetic ordering has been observed in epitaxially stabilized cubic CrN thin films\cite{Gall2002,Sanjines2002,Zhang2010}. The importance of the electron correlations in CrN has recently been shown experimentally\cite{Bhobe2010}. Most of theoretical calculations have considered only ordered magnetic structures\cite{Filippetti1999,Filippetti2000,Herwadkar2009} despite the fact that most experimental measurements are carried out above the Ne\'el temperature. 

In 2009, Rivadulla \textit{et al.} reported that the bulk modulus of CrN collapses at high pressures\cite{Rivadulla2009}, but their theoretical modeling neglected the interplay between the magnetism and structure on the local level as well as globally in the paramagnetic regime. Theoretical modeling taking disordered magnetism into account did not find any dramatic changes in compressibility\cite{Alling2010a}. Later experiments confirmed the absence of elastic softening upon the phase transition\cite{Wang2012}. In other words, above the critical Curie or Ne\'el temperature,there are examples of fully itinerant Stoner magnets but most magnetic systems and CrN in particular retain their magnetic moments even though the long range order between them is lost.
Thus when performing first-principle calculations, a disordered magnetic configuration must be considered to simulate and fully understand the physical properties of paramagnetic CrN at elevated temperatures\cite{Alling2010a}. It should be noted that magnetic materials above the magnetic transition temperature are not only spatially disordered but also dynamically disordered. Thus at higher temperatures the magnetic moments are changing randomly over time. Alling \textit{et al.}\cite{Alling2010} performed a detailed study on the effect of the magnetic disorder on the thermodynamics of paramagnetic CrN using disordered local moment method (DLM) implemented within two supercell frameworks, magnetic sampling method (MSM) and special quasi-random structure method (SQS). In their work, they obtained an orthorhombic to cubic phase  transition temperature as a function of pressure qualitatively in line with, but quantitatively higher than, the experimental measurements. It was suggested that the magnetic short-range order coupled to the vibrational degrees of freedom should also be of importance in order to determine the transition temperature in CrN. The importance of the vibrations was the main motivation behind Steneteg's work merging the disordered local moment model with \textit{ab initio} molecular dynamics (DLM-MD)\cite{Steneteg2012}. Recently, the effect of vibrations on the phase transition were investigated\cite{Hellman2013} by DLM-MD combined with temperature dependent effective potential (TDEP)\cite{Shulumba2014} method. Furthermore, Ti$_{1-x}$Cr$_x$N\cite{Alling2010b} and Cr$_{1-x}$Al$_x$N\cite{Alling2013} alloys have also been studied using these methods. Others have also investigated Cr$_{1-x}$TM$_x$N alloys using SQS approach\cite{Zhou2013} in which the Cr spin up and Cr spin down are randomly distributed in a disordered way in the supercell.    

The electronic structure of CrN is another controversial issue. Depending on the experimental circumstances, some groups have observed CrN to exhibit semiconducting behaviour in its AFM phase\cite{Browne1970}. This conclusion has been supportedsupported by some band structure calculations\cite{Filippetti1999,Filippetti2000} indicating that AFM CrN exhibits semiconducting behaviour whereas it behaves like a metal in its B1-rocksalt structure. This is in contradiction to measurments done by Herle \textit{et al.}\cite{herle1997} and more recent experiments on rocksalt PM CrN showing an increase in the resistivity with decreasing temperature, indicating that it is a semiconductor,\cite{Gall2002,Sanjines2002,Constantin2004} but the discussion is still open. 
   
This far almost all the theoretical investigations of CrN-based materials have assumed a stoichiometric nitrogen sublattice. This is in contrast to the experiments where the stoichiometry of TMN films can depend strongly on the specific choice of the synthesis technique and in the case of thin films, the choice of growth parameters. Therefore, the point defects are naturally present in the films. In particular, N vacancies ($V_N$), but also N interstitials ($I_N$) are believed to be important point defects existing in TMNs films grown by physical vapour deposition techniques \cite{Sundgren1985,Kang1999}. Several groups have experimentally studied how the electronic and mechanical properties of CrN change due to nitrogen stoichiometry\cite{Subramanian2012,Hultman2000,Hones2003,Greczynski2010a}. 

The nature of the conduction in CrN is affected by vacancies and interstitials as they contribute with carriers and also induce defect states in the band structure. In general the effect of defects and temperature-induced vibrations specially in paramagnetic systems remain unresolved. Due to the strong magneto-lattice interactions in PM systems and their tendency to deviate form the nominal 1:1 stoichiometry, the effects of the interaction of magnetic disorder, vibrations and structural relaxations are large. Thus more detailed and new methods are needed to study the behaviour of the vacancy, interstitial or substitutional atoms in paramagnetic materials. 

In this article, the impact of N vacancies ($V_{N}$) and interstitials ($I_{N}$) on the electronic structures and the defect formation energies in the PM cubic CrN at finite temperature are studied.


\section{Calculational details}
\label{2}
The formation energy of a single nitrogen defect is defined according to  
\begin{equation}
\label{eq:1}
E^{F} = E_{def} - E_{TMN} + \frac{x}{2} E_{N_2}
\end{equation}
in which $E_{def}$ is the total energy of the system in the presence of the defect, $E_{TMN}$ is the total energy of the host, defect free CrN system and $E_{N_2}$ is the energy of the $N_2$ molecule in the vacuum. $x$ can be either +1 for $V_N$ and -1 for $I_N$ defects.

In this work, all first-principles calculations are performed utilizing the projector augmented wave (PAW) method as implemented in Vienna \textit{Ab initio} Simulation Package (VASP)\cite{Kresse1996,Kresse1999,Kresse1993}. The Born-Oppenheimer molecular dynamics is used in our simulations. 
The paramagnetic state is approximated in our calculations through harnessing the disordered local moment model, introduced by Hubbard\cite{Hubbard1979,Hubbard1979a,Hubbard1981} and Hasegawa\cite{Hasegawa1979,Hasegawa1980} and applied within an electronic structure framework by Gyorffy\cite{Gyorffy1985}. Supercell implementations of DLM was investigated by Alling \textit{et al.}\cite{Alling2010} and the idea of combining disordered local moment model and molecular dynamics, DLM-MD, was introduced and implemented by Steneteg \textit{et al.}\cite{Steneteg2012} and has then be used to model mixing thermodynamics in $Cr_{1-x}Al_xN$\cite{Alling2013} and vibrational spectra of $CrN$\cite{Shulumba2014}. In this work we use both static supercell DLM and DLM-MD.  The electronic exchange-correlation effects are modeled via a combination of local density approximation with a Hubbard Coloumb term (LDA+U)\cite{Anisimov1991}. In this implementation of LDA+U, we have used the double-counting correction scheme suggested by Dudarev \textit{et al}\cite{Dudarev1998}. The value of the effective U ($U^{eff}=U-J$), shortly denoted as U here, is taken to be 3 eV for Cr $3d$ orbitals which is found to be the optimal value obtained theoretically from comparison of the structural and electronic properties of CrN with experimental measurements as reported in Ref\cite{Alling2010}. 

The energy cutoff for the expansion of the plane waves is 400 eV unless stated otherwise. Using Monkhorst-Pack scheme, the Brillouin zone is sampled on a $3\times3\times3$ grid containing 32 Cr and 32 N atoms when defects are not present. To obtain the density of states (DOS), a selection of configurations from MD steps are chosen and their electronic structure is recalculated with higher accuracy, that is with an energy cutoff of 500 eV and a grid of $5\times5\times5$ $k$ points. 

The calculations are also done for larger supercell of 216 atoms at T=0 K for the defect-free system and when the defects are introduced. Due to the expensive calculations, DLM-MD calculations for the large supercell are done only at two different temperatures to evaluate the vacancy formation energy. 



The simulations are carried out considering a canonical ensemble (NVT). In order to control the temperature of the simulation, the standard Nos\'e thermostat\cite{Nose1991} is used as implemented in VASP with the default Nos\'e mass set by the package.

The lattice constant of CrN is set to its calculated equilibrium zero temperature value, 4.129 $\AA$. It is worth to mention that we also ran several calculations using the experimental lattice constants ranging from 4.15 $\AA$ to 4.18 $\AA$ at specific temperatures, i.e. we cosider the effect of thermal expansion according to experiments.

In order to model the paramagnetic state in the static calculations, we have used the magnetic sampling method (MSM)\cite{Alling2010}. Within this approximation, the direction of the magnetic moments (up and down) of the Cr atoms for the 64-atom supercell are randomly distributed. A set of such samples with different magnetic configurations surrounding the defect and the constraint of a net total zero magnetization in the supercell, are then created. Each of these configurations are fully relaxed and their energies are calculated. The defect-free CrN system with different magnetic configurations is also sampled. All the configurations are relaxed and the average energy is calculated which gives us the term $E_{TMN}$ in eq. \ref{1}. The reported energy for static calculations in this paper is the average energy of these calculations which is chosen as the energy of the fictitious paramagnetic disordered state at 0 K\cite{Ponomareva2014}. It should be noted that the magnetically induced lattice relaxations are artificial consequences of the static magnetic order. However, the energies of these relaxations are likely to cancel if the same order is used in defect containing and pure CrN cell of eq. \ref{eq:1}. At least this was the case for the $Cr_{1-x}Al_xN$ in Ref. \cite{Alling2013} 

The DLM-MD model is introduced to simulate the paramagnetic materials at finite temperatures. In this method, the local magnetic moments are considered to be disordered and the magnetic state of the system is changed in a stochastic way with a given period during the MD simulation. In other words, the simulation starts with a random magnetic configuration and an MD simulation is run for some time during which the magnetic state is fixed. The magnetic configuration of the system is then substituted with another random set of moments, the MD is then continued and the process goes on until the given simulation time is finished. The time span between which the magnetic configuration is being kept fixed is called the "spin flip time", $\Delta t_{sf}$. During our simulations $\Delta t_{sf}$ is chosen to be 5 $fs$ which according to Ref.\cite{Steneteg2012} should practically correspond to adiabatic approximation with fast moments.  The Potential energy of the atomic system, basically the total energy minus the kinetic energy of the ions, is extracted from the DLM-MD calculations. The energy of the system, $E_{TMN}$ or in the case of defects $E_{def}$, is then obtained by averaging over the energies of DLM-MD time steps. Using eq. \ref{eq:1}, on can then calculate the formation energies of a specific temperature at which the simulation is carried out. We have to mention also that the temperature dependence of the $N_2$ molecule, $E_{N_2}$, in eq. \ref{eq:1} is not included. We have used the 0 K value of $E_{N_2}$ for obtaining the formation energies as our focus is to investigate if the lattice vibrations influence defect thermodynamics. The effect of gas phase thermodynamics is beside the topic of this work.

\section{Results}
\label{Three}
In the following we will describe in detail the results obtained for CrN including nitrogen vacancy and interstitial nitrogen. We first discuss the nitrogen vacancy case and then we consider the nitrogen interstitials. 

\subsection{N vacancy} 
\subsubsection{Static supercell DLM}
The first type of N defects corresponds to nitrogen vacancies. Relaxing the supercell containing only a single vacancy, we found that the N vacancy will produce an outward relaxation in the surrounding CrN matrix, in agreement with what is obtained by Jhi \textit{et al.}\cite{Jhi2001} and Tsetseris \textit{et al.}\cite{Tsetseris2007,Tsetseris2007a} for TiN, ZrN and HfN. 

In order to calculate the formation energy via magnetic sampling method in combination with special quasirandom structure method (MSM-SQS)\cite{Ponomareva2014}, the vacancy is moved around in a $2\times2\times2$ conventional unit cell with magnetic state set according to SQS-DLM method, placing it in different magnetic local environments. The average energy of MSM-SQS calculations is considered as $E_{def}$ term in eq. \ref{eq:1}. The calculated vacancy formation energy is 2.28 eV/supercell as reported in Table. \ref{table:1}. 

\subsubsection{DLM-MD}
\begin{figure}[!p] 
\begin{center} 
\includegraphics[width=\linewidth]{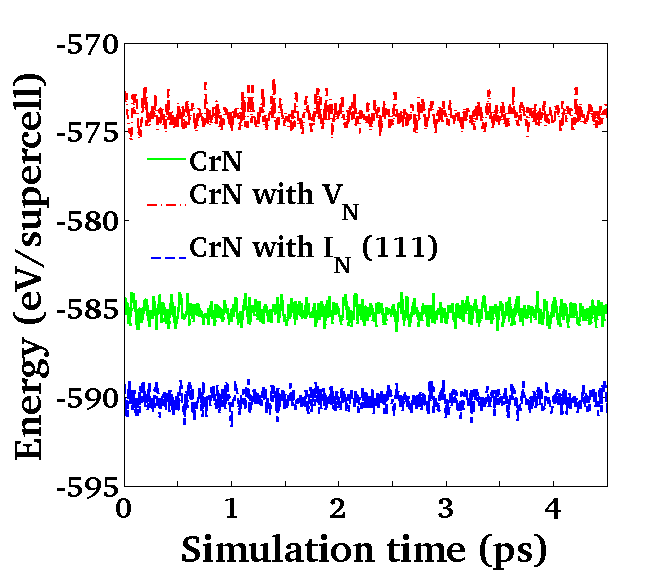} 
\caption{(Color online) Potential energy of paramagnetic CrN as a function of simulation time calculated at $T=300$ $K$ using DLM-MD method. The figure includes also the potential energy of cubic CrN in the presence of N vacancy and N interstitial ($I_N^{111}$) calculated at the same temperature.\label{fig:2}} 
\end{center} 
\end{figure}

The DLM-MD simulations are performed at different temperatures for a $k$-point grid of $3\times3\times3$. Fig.\ref{fig:2} shows the extracted potential energy from DLM-MD calculations at a specific temperature of T=300 K and as it is seen the potential energy is well conserved. The energy of the system, in the case of a defect $E_{def}$, is obtained by averaging over all MD energies (red dot-dashed line for vacancy and blue dashed line for interstitial). The energy, $E_{TMN}$ is also obtained from averaging over all the energies (green solid line) of CrN system in the absence of any defects. Obtaining the vacancy formation energies is then straightforward. The vacancy formation energies at 6 different temperatures are listed in Table \ref{table:1}. It is apparent that there is not a large change in vacancy formation energy as a function of temperature caused by vibrations. The given values show that $E^F$ is varying between 2.30 to 2.40 eV/supercell at various temperatures but in a random way indicating that it is a statistical noise and allowing us to estimate the error bar of 0.1 eV per defect in this method. And of course, the free energy of the defect should depend on temperature via the entropy contribution.  
Also the vibrational entropy is not explicitly considered here\cite{Shulumba2014}. 

The same calculations were performed for a larger defect free supercell of 216 atoms consisting of 108 Cr atoms and 108 N atoms and also for a supercell with a vacancy, 107 N atoms. The vacancy formation energy at T=300 K, in this case, is 2.35 eV/supercell. 

Comparing the vacancy formation energies obtained from static supercell DLM calculations and also from DLM-MD as listed in Table. \ref{table:1}, we can conclude that the DLM-MD calculations in this case justifies the static approximation. In particular, the good agreement between the static and DLM-MD calculations indicate that the artificial static relaxations present in the former method cancel each other, as they contribute almost equally to the the first and the second terms in the right-hand side of eq. \ref{eq:1}.  
   
\subsection{N Interstitial}

\subsubsection{Static supercell DLM}
\begin{figure}[!p] 
\begin{center} 
\includegraphics[width=\linewidth]{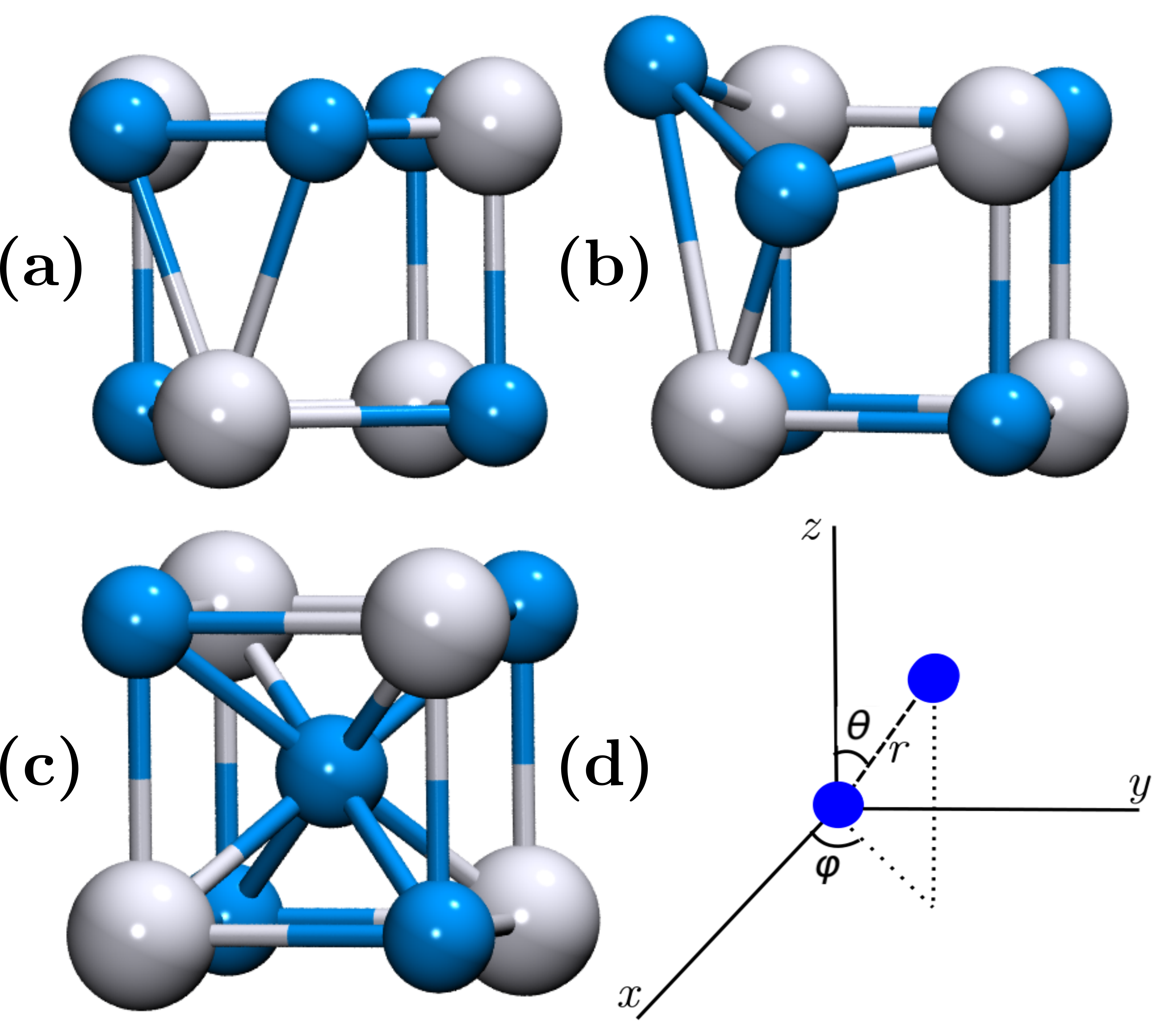} 
\caption{(Color online) Possible N interstitial initial positions in the CrN unit cell. The blue circles show nitrogen and the gray circles are Cr atoms. The figure demonstrates N interstitial in the (a) $I_N^{110}$ position (b) $I_N^{111}$ position and (c) $I_N^{tet}$ position. (d) The spherical coordinates related to the coordinates and lines depicted in Fig. \ref{fig:3}. \label{fig:1}} 
\end{center} 
\end{figure}

The second class of point defects that we consider are N interstitials. In contrast to the vacancy defects, different N-interstitial positions are possible. Probably the most straightforward consideration is an $I_N$ in a tetrahedral position, $I_N^{tet}$, shown in Fig. \ref{fig:1} (c). Other possible arrangements are the split-bond interstitials for which the excess N atom forms a N-N dimer with one of the nitrogen atoms in the CrN matrix centred at the N lattice position. This dimer can be aligned in different directions. Fig. \ref{fig:1} (a) and (b) demonstrate two different arrangements with the N-N bond pointing in \big<110\big> ($I_N^{110}$) and \big<111\big> ($I_N^{111}$) directions, respectively. 

For harnessing magnetic sampling method (MSM) in the case of interstitials, the excess nitrogen is kept in one position and the magnetic configuration of the Cr atoms surrounding it, is changed. The energy, $E_{def}$ is calculated in the same way as discussed earlier in this paper and the formation energies are obtained using eq. \ref{eq:1}. In this case, the defect-free CrN is also calculated for each magnetic sample. The relative stability results for both split-bond and tetrahedral interstitials is summarized in Table. \ref{table:1}. The formation energy of $I_N^{110}$ is 3.84 eV/defect which is 0.07 eV/defect higher in energy than the value obtained for $I_N^{111}$, 3.77 eV/defect. In the case of nitrogen interstitial in a tetrahedral position for a $2\times2\times2$ supercell consisting of 32 Cr atoms and 33 N atoms, we have done the calculations for 27 magnetic samples out of which 14 configurations relaxed to a split-bond $I_N^{111}$. The defect in 13 other magnetic samples remained in the tetrahedral position. This can be interpreted that $I_N^{tet}$ is at best a metastable site at T=0 K. 
The N-N dimer bond length in the case of $I_N^{111}$ is 1.27 $\AA$ which is larger than the $N_2$ molecule bond length, 1.1 $\AA$ as calculated with LDA. In order to understand this situation, we also studied the formation energy and the bond length variation of the N interstitials as a function of temperature.     
\subsubsection{DLM-MD} 
Fig.\ref{fig:2}, includes the potential energy of the system with a nitrogen interstitial in the split-bond $I_N^{111}$ configuration at T=300 K (the dashed blue line). The potential energy is well conserved as for the nitrogen vacancy case. $E_{def}$, is obtained by averaging over all DLM-MD energies. The formation energy is then calculated using eq. \ref{eq:1}.    

The data obtained for the relative stability of nitrogen interstitials show that the only stable configuration at elevated temperatures is the split-bond $I_N^{111}$. The $I_N^{111}$ formation energies at different temperatures are summarized in Table. \ref{table:1}. Unlike the vacancy case, the formation energy of the nitrogen interstitial exhibit an increasing trend as the temperature increases. It starts from 3.80 eV/supercell at T=300 K and rises to 3.97 eV/supercell at T=900 K except for the formation energy at T=1200 K which drops to 3.88 eV/supercell. The same argument about $E_{N_2}$ term in eq. \ref{eq:1}, mentioned in the beginning of Sec. \ref{Three}, applies also here. In our calculations, $E_{N_2}$ is considered to be temperature-independent and its 0 K value is used in all cases.   

\begin{figure*}[t]
\begin{minipage}[c]{0.47\linewidth}
\null
\includegraphics[width=\textwidth]{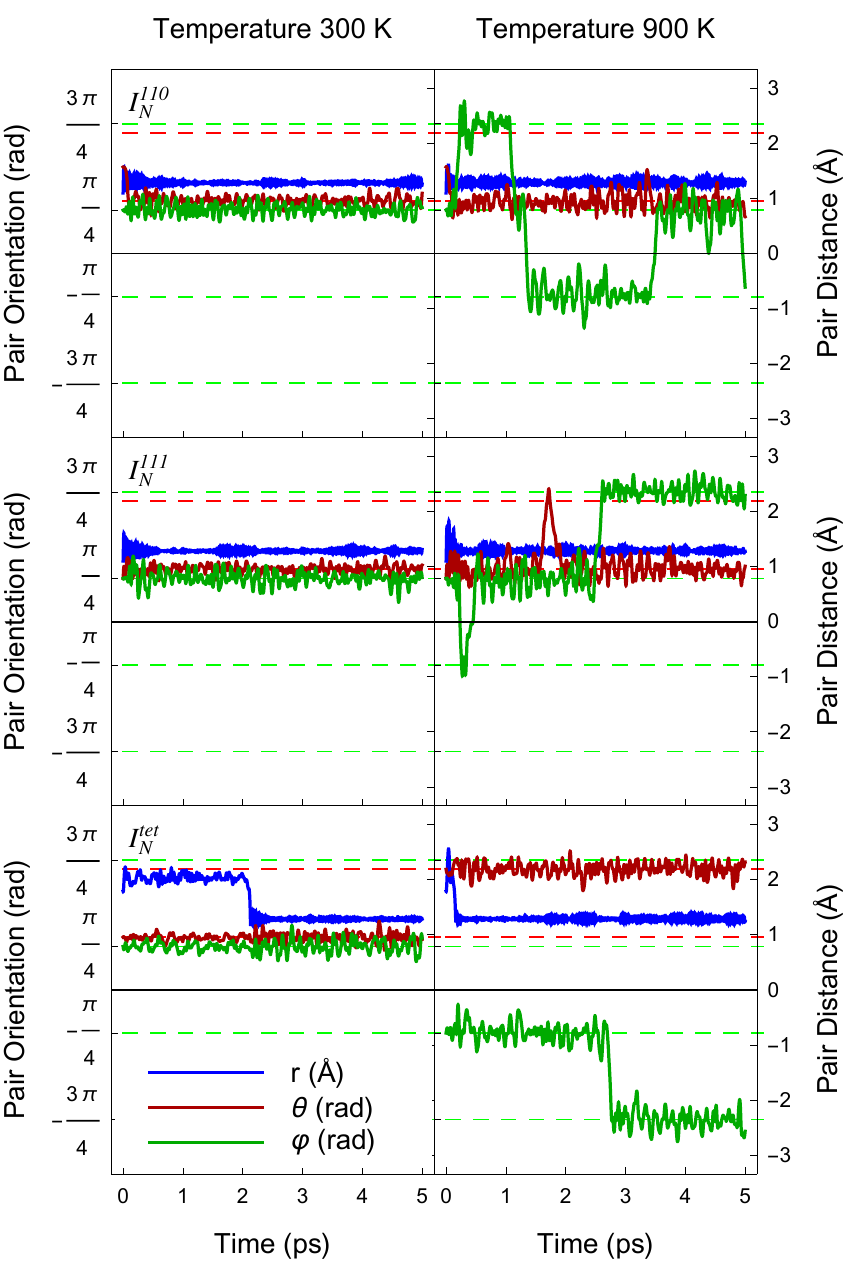} 
\end{minipage}\hfill
\begin{minipage}[c]{0.47\textwidth}
\caption{(Color online) Nitrogen pair (N($I_N$)-N(CrN)) dynamics in paramgnetic CrN as a function of simulation time at $T=300$ K (left column) and $T=900$ K (right column). Panels from top to bottom show split-bond \big<110\big>, split-bond \big<111\big> and tetrahedral configurations, respectively. The spherical coordinate system used in the Figure is shown in Fig. \ref{fig:1}. The blue color is the radial distance between the nitrogen pair in $\AA$. The angles show the orientation of the N-N bond vector and the dashed lines represent all the symmetry directions equivalent to \big<111\big> direction, \big<$\bar{1}$11\big>, \big<11$\bar{1}$\big> , \big<1$\bar{1}$1\big>, etc. The green color shows the azimuthal angle $\varphi$. The dashed green lines are different values of $\varphi$ between 0 and 2$\pi$. The red solid and dashed lines show the polar angle $\theta$ which varies between 0 and $\pi$. In the case of $I_N^{110}$, the value of $\theta$ (red solid line) should be constant $\pi/2$. However, as can be seen in the figure $\theta$ changes to $\pi/3$ and vibrates around this value indicating that the pair is in the \big<111\big> direction. \label{fig:3}}
\end{minipage}
\end{figure*}

In order to understand the relation between different interstitial positions, the nitrogen pair (the $I_N$ and the N from the CrN lattice) dynamics is studied in Fig. \ref{fig:3}. The figure depicts the $I_N$'s bond vector (N-N bond) coordinates in spherical coordinate system (see Fig. \ref{fig:1} (d)) as a function of simulation time. From Fig. \ref{fig:3} it can be seen that the interstitial nitrogen in the tetrahedral position at $T=300 K$ transforms into the split-bond after some time ($\sim 2 ps$), seen in the N-N distance $r$. Fig. \ref{fig:3} also demonstrates the same variables at $T=900 K$. It is clear from the figure that both $I_N^{110}$ and $I_N^{tet}$ configurations transform into  $I_N^{111}$ arrangement more or less immediately. The bond distance, $r$, is almost constant around 1.2 $\AA$ while some variations can be distinguished in the polar angle $\varphi$ in all three cases. The dashed lines show all the directions symmetry equivalent to \big<111\big> direction, namely \big<$\bar{1}$11\big>, \big<11$\bar{1}$\big>, \big<1$\bar{1}$1\big>, etc. At T=900 K, starting from angle $\pi/4$, green dotted line, the polar angle $\varphi$ goes almost immediately to $3\pi/4$ in the case of $I_N^{110}$ and oscillates around this value for a while, it then goes to $-\pi/4$,  remains there for some time and finally goes back to $\pi/4$. In the case of $I_N^{111}$, $\varphi$ oscillates around $\pi/4$ at the beginning and goes to $3\pi/4$ after almost 2.5 ps. For tetrahedral configuration, as mentioned the interstitial transforms to the $I_N^{111}$ position and the polar angle vibrates around $-\pi/4$ until 2.8 ps after which it goes to $-3\pi/4$ and oscillates around this angle during the rest of the simulation. The azimuthal angle $\theta$, on the other hand, is almost conserved and vibrates around the red dashed line showing $\pi/3$ in $I_N^{111}$ and around $2\pi/3$ in the tetrahedral case which is one of the equivalent directions to \big<111\big>, indicating that the tetrahedral N transforms to $I_N^{111}$. In $I_N^{110}$ case, $\theta$ should be $\pi/2$ and as it is seen in the Fig. \ref{fig:3}, the red solid line starts from $\pi/2$ but at both temperatures, it goes to $2\pi/3$ after a very short time. This indicates that the $I_N^{110}$ configuration is unstable. It transforms to $I_N^{111}$ configuration and the N-N dimer continues vibrating along \big<111\big> direction. The changes in the angles are basically jumps between the directions equivalent by symmetry to \big<111\big> direction. For further information regarding the dynamics see the supplementary documents.
 
\begin{figure*}[t]
\begin{minipage}[c]{0.47\linewidth}
\null
\includegraphics[width=\textwidth]{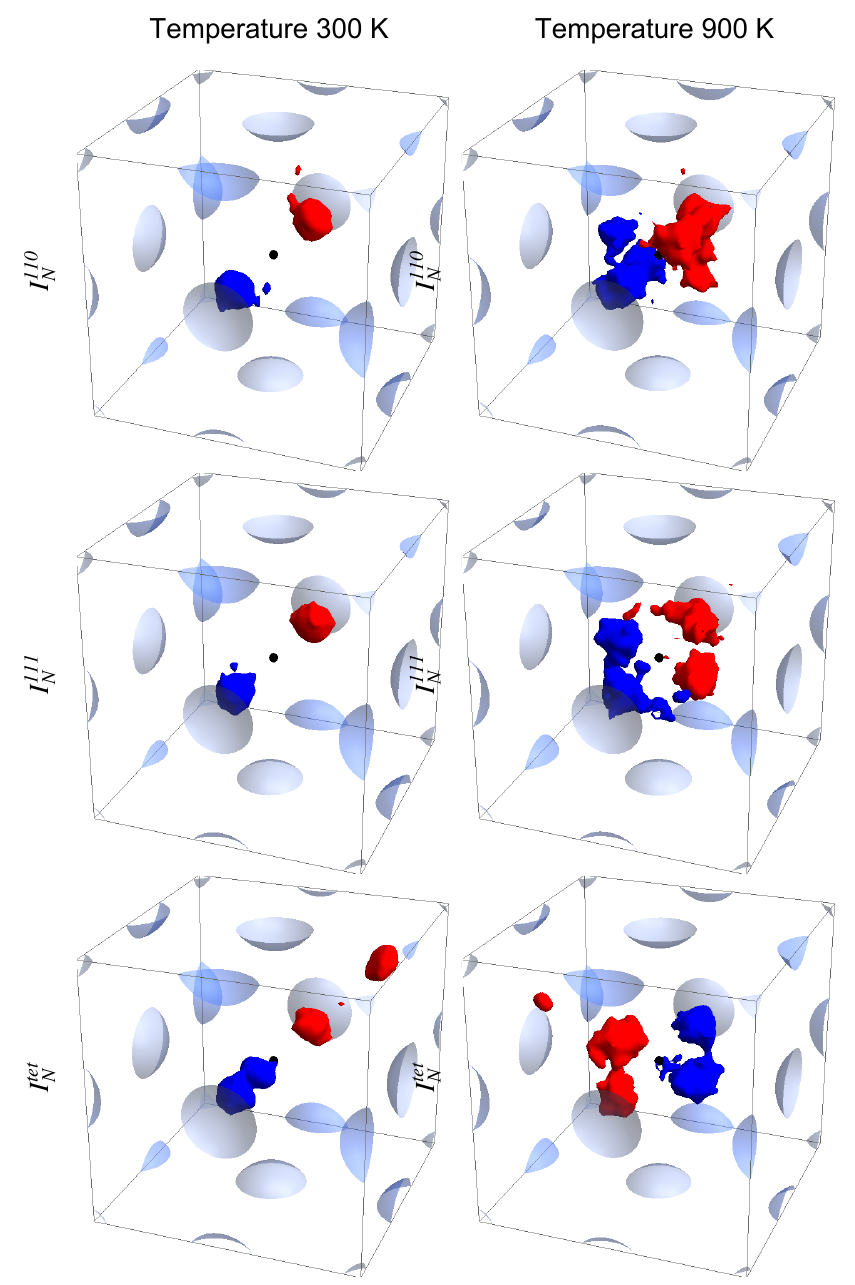}
\end{minipage}\hfill
\begin{minipage}[c]{0.47\textwidth}
\caption{(Color online) Nitrogen pair (N($I_N$)-N(CrN)) probability density in paramgnetic CrN at $T=300$ K (left column) and $T=900$ K (right column). Panels from top to bottom show split-bond \big<110\big>, split-bond \big<111\big> and tetrahedral geometries, respectively. Figure includes only the neighbouring Cr atoms (gray). Red shows the interstitial nitrogen and blue is the nitrogen from the CrN matrix. The starting geometry is shown by a black circle. The blue and red colors are densities that show where the two nitrogens in the pair spend most of their time during the simulation. The densities follow the trend shown also in Fig. \ref{fig:3}. For instance, the nitrogen in the tetrahedral configuration at T=300 transforms to $I_N^{111}$ position and it remains there as it can be seen the related panel in Fig. \ref{fig:3}. \label{fig:4}} 
\end{minipage}
\end{figure*}

To complete the picture, we also plotted the nitrogen pair position probability density to illustrate the spatial distribution of the N atoms during DLM-MD simulations. The red color shows the extra nitrogen added to the system and the blue color is the nitrogen from the CrN matrix pairing with the interstital N. Where the density is higher, indicates the position in which the nitrogen spends most of its time during the simulation. As can be seen in Fig. \ref{fig:4}, the tetrahedral case at T=300 K is in complete agreement with its counterpart in Fig. \ref{fig:3} in which the $I_N^{tet}$ transforms to $I_N^{111}$ position after some time. It is also visible that the $I_N^{110}$ moves to $I_N^{111}$ arrangement. The pair position probability density at $T=900 K$ is also shown in the right panel of Fig. \ref{fig:4}. 

The situation is similar for other temperatures except for the fact that the higher the temperature is, the more vibration is seen in the plots and the probability densities look spatially more spread.

Studying the N-N pair bond length shows that the average bond distance of the dimer varies between 1.28 $\AA$ to 1.3 \AA, depending on the temperature, which is in good agreement with the nitrogen pair distance reported in other TMNs\cite{Tsetseris2007a}. The obtained value is somewhat larger than both calculated (using LDA) and experimental value of the nitrogen molecule bond distance, 1.1 \AA. 

The movies obtained from DLM-MD calculations (see the supplementary documents) show that the N-N bond wants to have a particular orientation (\big<111\big> direction). This is what is seen in Fig. \ref{fig:3}. This actually indicates that the N-N dimer is not freely rotating in any direction. This implies that the nitrogen pair is in fact bonded to the CrN matrix. A free nitrogen molecule exhibits a triple bond but according to the fact that the N-N bond is streched and larger than 1.1 $\AA$ for N-interstitial in CrN,  we can argue that the nitrogen dimer is better described to have an internal double bond with two electrons, which can freely interact with the CrN lattice.  

To critically evaluate this suggestion, the electronic structure study is needed. It is discussed in the next section.    

\section{Electronic Structure}
\begin{figure}[!p] 
\begin{center} 
\includegraphics[width=\linewidth]{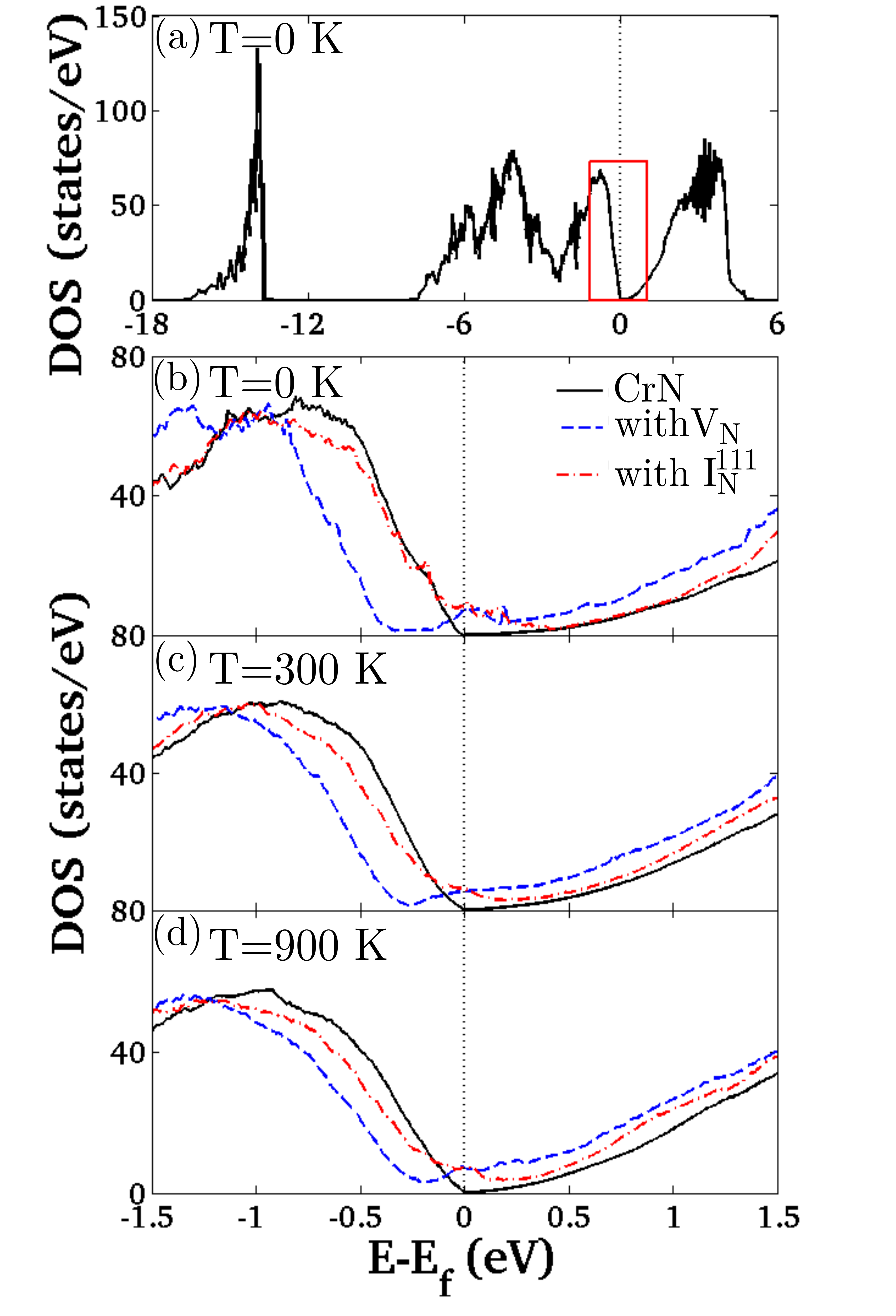} 
\caption{(a) Total density of states of CrN at T=0 K obtained from static supercell DLM calculations. In order to have a detailed view of the region around the Fermi level, in panels (b-d) we show only the area marked with a red rectangle in Panel (a). (b) DOS obtained in static calculations. In the two bottom panels we show the total density of states of CrN obtained from DLM-MD calculations at temperatures (c) 300 K and (d) 900 K. The results are based on a 64-atom supercell with 32 Cr atoms and 31 (vacancy) or 32 (defect-free) or 33 (interstitial) N atoms. The Fermi level, the vertical dashed line, is aligned to zero. \label{fig:5}} 
\end{center} 
\end{figure}

We have studied the effect of the point defects on the electronic properties of cubic paramagnetic CrN. Fig. \ref{fig:5} shows the density of states (DOS) of defect-free CrN obtained in static calculations using our approach along with the DOS of the system in the presence of point defects. The supercell consists of 32 Cr atoms and 32 N atoms in the defect-free case and 31 or 33 N atoms when vacancy or $I_N^{111}$ interstital is present. We have also calculated the density of states for a large supercell consisting of 216 atoms, 108 Cr and 108 N and for the case of N vacancy with 107 N atoms (not shown here). Defects could add or remove electrons to the system which will be visible as a shift to the Fermi level with respect to the bands in a defect-free supercell. They will affect the conductivity of the material. Such effects are apparent in Fig. \ref{fig:5} where there is a small gap at the Fermi level in the absence of nitrogen defects, in agreement with the previous studies\cite{Alling2010}. The presence of a defect results in the closing of this very narrow gap. 
As can be seen in the figure, Fig. \ref{fig:5} (blue dashed line), nitrogen vacancy introduces a peak at 
around 0.2 eV. The peak primary corresponds to Cr spin down non-bonding states (above $E_F$) with a small admixture of $N_p$ states. When a vacancy is present in the system, 3 electrons are released from the bonds between the neighbouring Cr atoms and the nitrogen. They will increase the band filling which will lift the Fermi level towards the conduction band. 

In the case of nitrogen interstitial, one can assume a CrN system with a vacancy in which the $V_N$ is substituted with a nitrogen dimer instead. Assuming that the N-N dimer exhibits a double bond, as we discussed in previous section, it leaves two electrons to bond with its Cr neighbours. The remaining one electron will increase the band filling as compared to defect-free CrN. Thus adding a N-interstitial will add one electron to the system. This is in agreement with the obtained results and can explain why the DOS in the case of N interstitial (red dashed dotted line in Fig. \ref{fig:5}) is shifted just about to the one-third of the shift one sees in the vacancy case (blue dashed line in Fig. \ref{fig:5}) from its position in the ideal CrN (solid black line in Fig. \ref{fig:5}) density of states. The apparent shift of the Fermi level towards the valence band, judged by the position of the DOS minimum is just an artefact from the position of the defect state. 
Two peaks are visible around the Fermi energy at 0.02 and -0.07 eV. The peak at 0.02 eV corresponds to combination of spin up and spin down Cr non-bonding states. The later, -0.07 eV, pertains to the combination of Cr spin up and down non-bonding (below $E_F$) states. Both peaks exhibit some $N_p$ character as well.   

The aim of the paper is to study the effect of the defects on the electronic properties of paramagentic cubic CrN at elevated temperatures. To obtain the density of states in this case, from DLM-MD calculations, we randomly extracted several MD samples. Then we calculated the density of state for each of them using a $5\times 5\times 5$ $k$-grid and an energy cut-off of 500 eV. The total DOS is then derived by averaging over all these calculations.

Fig. \ref{fig:5} (c) and (d) show the total density of states from the DLM-MD calculations at two different temperatures, 300 K and 900 K, respectively. At both temperatures, as it is seen, the gap has vanished due to the presence of defects. For the case of pure CrN DOS (black solid line in Fig. \ref{fig:5}), it is difficult to draw a conclusion from our calculations at what exact temperature the band gap closes. We can argue that the band gap of defect-free CrN is smeared out due to vibrations. However, at 900 K the DOS at the Fermi level is clearly non-zero. At 300 K our results indicate a boarder case were the band-gap is about to close. The density of states around the Fermi level, follows quite the same trend as in the static zero temperature case, Fig. \ref{fig:5} (b), meaning that the nitrogen vacancy moves the Fermi level towards conduction band. The same explanation applies for the N interstitial case as discussed in the static case. The peaks which were present in the density of states at zero temperature, in the vacancy case, have vanished at higher temperatures, i.e, that the temperature broadens the fine structure of the DOS at T=0 K.


\section{Conclusion}


We have applied a recently introduced model, DLM-MD, to study the formation energy, geometry and electronic structure of the nitrogen defects in paramagnetic CrN as a function of temperature. We have calculated the formation energies of both nitrogen vacancy and interstitials. The vacancy formation energy does not show a strong dependence on temperature. However, the authors would like to underline that the role of the temperature in the $N_2$ gas free energy is not considered in these set of calculations. 
The only stable N interstitial according to our calculations is the split-bond in the $\big<111\big>$ direction. The other two considered configurations transform to this position after at most a few picoseconds of simulation at finite temperature. Note that the formation energy for $I_N^{111}$ increases as the temperature increases.

Detailed studies of the dynamics of N interstitials at elevated temperatures show that the $I_N$ bonds to a nitrogen from the CrN matrix. The bond is stable and the N-N dimer oscillates around its center of mass, keeping its position. In other words, the N-N can rotate between different symmetry equivalent $\big<111\big>$ directions. 

In the case of N vacancy formation energy, there is a good agreement between the SC calculations and DLM-MD. Thus, the self-consistent calculations can be trusted and sufficiently enough to decide if the structure is stable. However, for N interstitals, the SC calculations cannot be considered enough to check the stability of the defect. As we have observed during our theoretical studies, for instance, the tetrahedral configuration can exist if we limit ourselves to the static approximation while DLM-MD calculations reveals the instability for a defect in this position at room temperature and above.   

Electronic structure calculations show that the ideal CrN exhibit a small band gap indicating a semiconducting behaviour in its cubic PM phase. The presence of N point defects, closes this gap, resulting in density of states which show metallic characteristics. Both N vacancy and N interstitial act as donors with $V_N$ donating 3 electrons and $I_N^{111}$ donating 1 electron to the system. 
 
\section{Acknowledgements}
Th authors would like to thank Swedish National Infrastructure for Computing (SNIC) at NSC center for providing the computational resources. The financial support from Swedish Research Council (VR) through grant No. 621-2011-4426 is gratefully acknowledged. B. A. acknowledges the financial support also from VR, grant No. 621-2011-4417. I. A. A. acknowledges the support from the Grant of Russian Federation Ministry for Science and Education (grant no. 14.Y26.31.0005). We would also like to assert our appreciation to Prof. Lars Ojam\"ae for useful discussions.

\bibliography{/home/elin/Documents/library/Paper4-ref.bib}

\begin{table*}[ht!]\centering
\footnotesize 
\begin{threeparttable}
\begin{ruledtabular}
\caption{\label{table:1} Relative stability of nitrogen point defects in paramagnetic CrN obtained from Static DLM and DLM-MD calculations at six different temperatures.}
 \begin{tabular}{ cc|c|c|c|c|c|c|c}
  \multirow{2}{*} {  Defect } & \multirow{2}{*} { Geometry } & \multirow{2}{*} {$E^{F}$ $^{1}$(eV/defect) \textbf{Static DLM}} & \multicolumn{6}{ c }{$E^{F}$(eV/defect) \textbf{DLM-MD} } \\ \cline{4-9} 
                   &                  &                  & 300 K & 600 K & 700 K & 900 K& 1000 K& 1200 K\\
  \hline \hline
$V_N$ & --- & 2.28 & 2.37 & 2.40 & 2.38 & 2.32 & 2.37 & 2.30\\
$I_N^{tet}$ & Tetrahedral & Metastable $^{2}$ & ---$^{3}$ & --- & --- & --- & --- & ---  \\
$I_N^{110}$ & Split-bond \big<110\big>  & 3.84 & ---$^{3}$ & --- & --- & --- & --- & --- \\
$I_N^{111}$ & Split-bond \big<111\big> &  3.77 & 3.80 & 3.84 & 3.87 & 3.87 & 3.97 & 3.88 \\   
 \end{tabular}
\end{ruledtabular}
\begin{tablenotes}
      \footnotesize
      \item $^{1}$ The energies are obtained from MSM.
      \item $^{2}$ During our MSM calculations we noticed that an interstitial in the tetrahedral configuration would either remains in the tetrahedral position or go into the split-bond $I_N^{111}$ configuration (See the text).
      \item $^{3}$ DLM-MD calculations at various temperatures show that the tetrahedral and the split-bond \big<110\big> configurations are unstable. Instead the N-N pair positions itself along the \big<111\big> direction. 
    \end{tablenotes}
  \end{threeparttable}
\end{table*}

\end{document}